\documentclass[12pt]{article}
\usepackage{eurosym}
\usepackage{graphicx}
\usepackage[utf8]{inputenc}
\usepackage[T1]{fontenc}
\usepackage{indentfirst}
\usepackage[margin=0.8in,nomarginpar]{geometry}
\usepackage[final]{hyperref}
\usepackage{amsmath}
\usepackage{hyperref}
\usepackage{cite}
\usepackage{subcaption}
\usepackage{caption}
\usepackage{amssymb}
\usepackage{multirow}
\usepackage{color}
\newcommand{\eps}{\epsilon}

\newcommand{\be}{\begin{equation}}
\newcommand{\ee}{\end{equation}}
\newcommand{\ba}{\begin{array}}
\newcommand{\ea}{\end{array}}

\hypersetup{
colorlinks=true,
linkcolor=blue,
citecolor=blue,
filecolor=magenta,
urlcolor=blue
}

\begin{document}

\title{%On The Wigner Parameter in the Dunk Formalism for Harmonically Trapped Ideal Bose Gases
%%%%%%%%%%%%%%%%%%%%%%%%%%%%
%Analysis of the Wigner Parameter within Dunk Formalism for Harmonically Trapped Bose Gases
Bounding the Wigner Deformation Parameter in Harmonically Trapped Bose Gases
%
% Investigating the Wigner Parameter in the Dunk Formalism \\ for Ideal Bose Gases in Harmonic Traps
}
\author{M. Benarous\thanks{%
m.benarous@univ-chlef.dz} \\
%EndAName
Laboratory for Theoretical Physics and Material Physics, Faculty of Exact \\
Sciences and Informatics, Hassiba Benbouali University of Chlef, Algeria.
\and A. Hocine%
\thanks{%
ah.hocine@univ-chlef.dz} \\
%EndAName
Laboratory for Theoretical Physics and Material Physics, Faculty of Exact \\
Sciences and Informatics, Hassiba Benbouali University of Chlef, Algeria. 
%\and F. Merabtine%
%\thanks{%
%f.merabtine@univ-chlef.dz} \\
%EndAName
%Laboratory for Theoretical Physics and Material Physics Faculty of Exact \\
%Sciences and Informatics, Hassiba Benbouali University of Chlef, Algeria.
\and B. C. L\"{u}tf\"{u}o\u{g}lu\thanks{%
bekir.lutfuoglu@uhk.cz (corresponding author) } \\
%EndAName
Department of Physics, Faculty of Science, University of Hradec Kralove, \\
Rokitanskeho 62/26, Hradec Kralove, 500 03, Czech Republic.
\and B. Hamil\thanks{%
hamilbilel@gmail.com} \\
%EndAName
Laboratoire de Physique Math\'{e}matique et Subatomique,\\
Facult\'{e} des Sciences Exactes, Universit\'{e} Constantine 1, Constantine,
Algeria. 
}
\date{}
\maketitle

\begin{abstract}
By examining the internal energy and the heat capacity of a harmonically trapped ideal Bose gas within the Dunkl formalism, we show that the Wigner parameter influences the slopes of these thermodynamic functions in the critical region, reflecting its role in modifying the statistical properties of the system. However, despite these modifications, the phase transition itself retains the same order and critical exponents as in the standard case, in accordance with symmetry arguments. Furthermore, upon analyzing the classical behavior, we establish both upper and lower bounds for the Wigner parameter by ensuring thermodynamic consistency in different temperature regimes.
\end{abstract}

\newpage
\section{Introduction}

Symmetry principles are fundamental in shaping the laws of physics, governing interactions, conservation laws, and the structural properties of quantum systems. Among these, reflection symmetry is particularly significant, as it determines wave function behavior, shapes selection rules, and influences spectral properties across diverse physical contexts. In systems ranging from condensed matter to quantum statistical ensembles, the presence or breaking of reflection symmetry can lead to profound consequences, including phase transitions, topological phenomena, and modifications to the energy spectrum which in turn impact thermodynamical features. These effects are especially pertinent in statistical mechanics, where symmetry considerations influence partition and correlation functions, directly affecting macroscopic observables.

In the literature, a powerful mathematical framework for incorporating reflection symmetry in quantum systems is provided by the Dunkl formalism, which introduces the Dunkl derivative~\cite{Dunkl1}:
\begin{equation}
D = \frac{\partial}{\partial x} + \frac{\theta}{x} \Big(1-\hat{R}\Big). \label{dunkl}
\end{equation}
This operator generalizes the classical differential operator by incorporating the reflection operator $\hat{R}$, which is associated with Coxeter groups~\cite{Dunkl1}:
\begin{equation}
\hat{R} = \left(-1\right)^{x\frac{\partial}{\partial x}}, \qquad \text{where} \qquad \hat{R} f(x) = f(-x).
\end{equation}
The parameter $\theta$, sometimes referred to as the Wigner parameter~\cite{phys2, Chung2021}, is named after Wigner’s seminal work~\cite{Wigner1950}. In this formalism, the Dunkl derivative replaces the ordinary spatial derivative
\begin{equation}
\frac{\partial}{\partial x} \longrightarrow D,
\end{equation}
modifying the underlying symmetry structure of the system. Following~\cite{Dunkl1}, the Dunkl formalism has been widely applied across various areas of theoretical physics, particularly in systems exhibiting symmetries and quantum many-body problems. It has played a crucial role in studying deformed Heisenberg algebras~\cite{Plyushchay1994, Plyushchay1996, Miky1}, classical root systems~\cite{Hikami1996}, and the incorporation of reflection symmetry in quantum mechanics~\cite{Plyushchay1997, Miky2, Miky3}. Furthermore, it has been instrumental in investigations of bosonized supersymmetry~\cite{Gamboa1999, Plyushchay2000, Horvathy2010}, fractional helicity and Lorentz symmetry breaking~\cite{Klishevich2001}, and anyon wave equations~\cite{Horvathy2004}. The formalism has also been employed in studies of the Calogero-Sutherland models~\cite{Kakei1996, Lapointe1996} and the hydrogen atom via the Wigner–Heisenberg algebra~\cite{Rodrigues2009}.  

More recently, quantum systems with reflection symmetry have gained increasing attention due to their significance in diverse fields such as condensed matter physics~\cite{Bilel5}, integrable systems~\cite{Genest20131, Genest20132, Genest20141, Isaac2016, Ghazouani2020, Najafizade20223}, and quantum information theory~\cite{Debraj2024}. The Dunkl formalism has also been employed in statistical mechanics~\cite{HH2021, phys5, Bilel3, HO24, phys6} and the study of stationary oscillators, both non-relativistic~\cite{Genest20142, Salazar2017, Mota2022s, Halberg2022, Samira20222, Dong2023} and relativistic~\cite{Sargol2018, Mota20181, Ojeda2020, Mota20211, Mota20212, Bilel20221, Merad2021, Merad2022, Askari2023}, including their thermal properties~\cite{Ubriaco2014, Dong2021, Bilel20222, Rouabhia2023}. Additionally, it has been extended to time-dependent~\cite{Benchikha20241,  Lutfuoglu2025} and angular-dependent oscillators, Coulomb potential problems~\cite{Vincent3, Salazar2018, Ghazouani2019, Ghazouani2021}, and quantum systems in noncommutative~\cite{Samira2022} and curved spacetimes~\cite{Najafizade20221, Najafizade20222, Ballesteros2023}. Other recent developments include applications in broader quantum mechanical frameworks \cite{Junker2023, Quesne2024, Bougerne, Bouguerne20243, Junker20241, Benzair2024, Benchikha20242, Schulze20243, Schulze20244, Schulze20245, Mota20241, Mota20242, Hamil20251, Benchikha2025}. 

In most of these studies, the free Wigner deformation parameter naturally emerges in parity-dependent solutions, making the energy spectrum explicitly dependent on an arbitrary parameter. It has been suggested that the Wigner parameter can serve as a fitting parameter, providing a bridge between experimental results and theoretical predictions. Typically, authors have demonstrated that the Wigner parameter must be constrained from below, satisfying the condition $\theta > -1/2$ for the formalism to remain valid. However, a critical question remains: does an upper limit for the deformation parameter exist? This unresolved issue poses significant challenges, as the absence of a well-defined upper bound for the Wigner deformation parameter could lead to unphysical predictions.

Indeed, Ubriaco \cite{Ubriaco2014} demonstrated that, for a collection of $N$ bosons, the "free" deformed Hamiltonian can be expressed as  
\begin{equation}  
H= \sum_i \epsilon_i\tilde{N}_i,  
\end{equation}  
where $\epsilon_i$ is the energy of the \textit{i}-th level, and $\tilde{N}_i$ is the deformed number operator, satisfying  
\begin{equation}  
\tilde{N}_i |n\rangle=  
\begin{cases}  
n_i |n\rangle, & \text{if } n_i \text{ is even}, \\  
(n_i+2\theta) |n\rangle, & \text{if } n_i \text{ is odd},  
\end{cases}  
\end{equation}  
where $|n\rangle=|n_1,n_2,n_3...\rangle$ represents the occupation basis. It follows that if \( \theta \) is unbounded, the total number of particles is also unbounded.  

Furthermore, by applying the inverse transformation corresponding to Eq. (\ref{dunkl}), the Hamiltonian can be rewritten in terms of the "original" creation and annihilation operators as  
\begin{equation}  
H= \sum_i \epsilon_i \left[N_i+\theta\left(1-(-1)^{N_i}\right)\right],  
\end{equation}  
where $N_i$ is the usual number operator. This expression explicitly demonstrates that an unrestricted Wigner parameter leads to an unbounded energy. More significantly, it reveals that a non-interacting system within the Dunkl formalism corresponds to an interacting system in the standard framework (and vice versa), with the coupling constant dictated by \( \theta \). Consequently, the application of the Dunkl formalism to quantum gases provides a rigorous mathematical framework for describing particle interactions through deformed ideal systems.

In a recent paper \cite{HO24}, we explored harmonically trapped ideal Bose gases within the Dunkl formalism. Our findings revealed that these gases exhibit behavior distinct from the non-deformed case, both in the degenerate regime and the classical limit. The analysis was carried out with a free Wigner parameter, and while we highlighted several interesting properties that diverge from the standard case, we did not focus extensively on the Wigner parameter itself, which is responsible for these differences.

The goal of this work is to further test the Dunkl formalism by extending the results from \cite{HO24} to the \( d \)-dimensional case. This will provide new insights into the quantum dynamics of trapped systems and offer exact solutions to problems in various geometries. A secondary objective is to explore additional constraints on the Wigner parameter by examining both the critical and classical regimes, thus improving the framework's applicability and predictive power. Lastly, considering the formal analogy of these systems with homogeneous gases, we will demonstrate that the final results also apply to untrapped ideal Bose gases.

The paper is organized as follows. In Section \ref{sec2}, we extend the main results of our recent work \cite{HO24} to the $d-$dimensional case.  In Section \ref{sec3}, we discuss the degenerate regime, and we demonstrate that the Bose-Einstein transition is only possible when $\theta$ obeys the aforementioned consistency condition. Furthermore, we examine the discontinuity of the heat capacity in three dimensions in terms of the Wigner parameter and show that it tends to a saturation value for strong deformations. In the two-dimensional case, we show that the transition is continuous, resembling the BKT transition, but with the heat capacity exhibiting a sharp peak at the transition. However, this peak increases as \( \theta \) decreases. Section \ref{sec4} examines the high-temperature regime, where we show that the thermodynamic functions display distinct classical behaviors depending on the value of \( \theta \). This enables us to establish an upper bound for the Wigner parameter to maintain the correct classical behavior. Finally, we conclude by summarizing our results and providing perspectives for further exploration of the Wigner parameter's role.

\section{Harmonically trapped Ideal Dunkl-Bose gas in d-dimensions} \label{sec2}

In this section, we extend the results of \cite{HO24} to a $ d $-dimensional system. Before proceeding, it is essential to highlight that, within a deformed formalism, two equivalent approaches may be employed for the derivation of thermodynamic functions \cite{hamil2022}:  
\begin{itemize}  
    \item One approach involves using the deformed commutation relations while retaining the standard Hamiltonian. In this case, deformed integration measures must be introduced, leading to a modification of the density of states \cite{chang2002}.  
    \item Alternatively, one may adopt the standard commutation relations with a modified Hamiltonian, which results in a deformation of the occupation number operator \cite{Ubriaco2014, phys5, bosso2018}.  
\end{itemize}  
In the following, we proceed with the latter approach. 

We begin by considering an ideal Bose gas composed of $ N $ neutral atoms, each with mass $ m $, confined within a $ d $-dimensional harmonic potential with a mean frequency $ \omega $. The deformed Hamiltonian of the system is given by:  
\begin{equation}  
H=-\frac{\hbar^2}{2m}D^2+\frac{1}{2}m\omega^2 x^2,  
\label{ham}  
\end{equation}  
where $ D $ represents the $ d $-dimensional extension of the Dunkl derivative (\ref{dunkl}), and $ x $ is the $ d $-dimensional position vector.  

We now present a few remarks on the Hamiltonian. First, for simplicity, we employ a single Wigner parameter instead of $d$. Furthermore, the Hamiltonian \eqref{ham} clearly preserves the symmetries of its non-deformed counterpart. As a result, the order of the phase transition and the associated critical exponents are expected to remain unchanged under the deformation.

By considering the grand canonical density operator 
\begin{equation}
\exp{(-\beta (H - \mu N))},    
\end{equation}
where $ \beta = 1/(k_B T) $ and $ \mu $ is the chemical potential of the gas, it can be readily shown \cite{Ubriaco2014, phys5} that the deformed partition function takes the form  
\begin{equation}
Z=\prod\limits_{i=1}^{\infty}
\frac{1+e^{-\beta \left( 1+2\theta \right) \left(
\epsilon _{i}-\mu \right) }}{1-e^{-2\beta \left( \epsilon _{i}-\mu
\right) }}.  \label{pf}
\end{equation}  
It is evident that in the limit $ \theta \to 0 $, this expression reduces to the well-known partition function of an ideal Bose gas. Here, $ \epsilon_i $ denotes the single-particle eigenenergy of state "$i$".  

The total particle number and internal energy can be obtained using the standard thermodynamic relations:  
\begin{equation}
N=\frac{1}{\beta }\frac{\partial }{\partial \mu }\log Z\bigg|_{T,V},  
\qquad  
U=-\frac{\partial \log Z}{\partial \beta}\bigg|_{z,V},  \label{UN}
\end{equation}  
which yield  
\begin{equation}
N=\sum_{i=0}^{\infty}n_{\theta}(\epsilon_i), \qquad
U=\sum_{i=0}^{\infty}\epsilon_i n_{\theta}(\epsilon_i),
\end{equation}  
where  
\begin{equation}
n_{\theta}(\epsilon_i)=\frac{2}{e^{2\beta\epsilon _{i}}z^{-2}-1}+
\frac{1+2\theta}{e^{\beta (1+2\theta )\varepsilon _{i}}z^{-(1+2\theta )}+1}.
\label{distrib}
\end{equation}  
The function $ n_{\theta} $ represents the occupation probability of state "$i$", while 
\begin{equation}
  z = \exp{(\beta \mu)},  
\end{equation}
denotes the fugacity of the gas. Notably, $ n_{\theta} $ corresponds to a deformed Bose-Einstein distribution, which reduces to the standard form when $ \theta = 0 $. Within this formulation of the Wigner formalism, the occupation probabilities are modified in a manner reminiscent of $ \kappa $-statistics \cite{kanidakis2003} and $ q $-statistics \cite{wang2000}. However, a key distinction is that, unlike these alternative approaches, the deformation in this case arises from a modification of the Heisenberg algebra.

Let $ N_0 = n_{\theta}(0) $ represent the number of atoms in the ground state, and $ N_{\epsilon} $ the number of atoms in the excited states. We then obtain:
\begin{equation}
N_{\epsilon}=\sum_{i\ne 0}^{\infty}\left(\frac{2}{z^{-2} e^{2\beta\epsilon _{i}}-1}+%
\frac{1+2\theta}{z^{-(1+2\theta ) e^{\beta (1+2\theta )\epsilon _{i}}}+1%
}\right) ,  \label{NEX}
\end{equation}
which can be computed using the exact spectrum of the harmonic oscillator. Alternatively, a simpler approach is to assume a continuous spectrum 
\begin{equation}
  \epsilon(x,p) = \frac{p^2}{2m} + \frac{1}{2}m\omega^2 x^2,  
\end{equation}
and introduce the density of states,
\begin{equation}
\rho (\epsilon)=
\int \frac{d^dp\,d^dx}{(2\pi\hbar)^{d/2}}
\delta(\epsilon-\epsilon(x,p))=
\frac{\epsilon^{d-1}}{(\hbar\omega)^d \Gamma(d)}, \label{rho}
\end{equation}
where $ \Gamma(d) $ is the Euler gamma function. This semi-classical approximation significantly simplifies the calculations and is highly accurate for large $ N $, which we will focus on henceforth \cite{gro1}. 

By using the density of states (\ref{rho}), we obtain the following expressions:
\begin{eqnarray}
N_{\eps}&=&t^d g_d(z,\theta) \label{s},\\  
U/\hbar\omega&=&d\, t^{d+1} g_{d+1}(z,\theta),
\label{uu}
\end{eqnarray}
where $ t = \frac{1}{\beta \hbar \omega} $ is the reduced temperature, and $ g_d(z, \theta) $ is the generalized Bose function given by:
\begin{equation}
g_d(z,\theta) =g_d(z)+g_d(-z)-\frac{1}{\left( 1+2\theta \right) ^{d-1}}g_d(-z^{1+2\theta}) , \label{GS}
\end{equation}
which is defined as a combination of the standard Bose (Polylogarithmic) functions $g_d(z)$:
\begin{equation}
g_d(z)=\frac{1}{\Gamma (d)}\int_{0}^{\infty }\frac{x^{d-1}}{z^{-1}e^{x}-1}dx.
\end{equation}
It is important to note that for $\theta=0$, Eq. \eqref{GS} reduces to  
\begin{equation}
g_d(z, 0) = g_d(z).    
\end{equation}
However, as we will demonstrate later, significant differences arise when $ \theta \neq 0 $, particularly in the critical region ($ z \to 1 $) and the classical regime ($ z \to 0 $), due to the distinct properties of these functions.

The transition temperature is traditionally defined as the point where $N_{0} \simeq 0$ (implying $N_{\eps}\simeq N$) and $z \simeq 1$. For this condition to hold, $N_{\eps}$ and, consequently, $g_d(z,\theta)$ must remain bounded, which imposes the constraint $\theta>-1/2$  for any value of $z$. Notably, this constraint complements the previously established lower bound for consistency. With this understanding, and applying (\ref{s}), we can readily determine the $d$-dimensional case condensation temperature within the Dunkl formalism as:
\begin{equation}
t_c(\theta)=\left[ \frac{N}{g_d(1,\theta )}\right] ^{1/d}. 
\label{Tc}
\end{equation}
Here, one can recover the standard Bose-Einstein Condensation (BEC) temperature for $ \theta = 0 $. Moreover, since $ g_d(1, \theta \to -1/2) \to \infty $, the critical temperature approaches zero, indicating the absence of a BEC transition in this limit. Additionally, as $ d \to 1 $, $ g_d(1, \theta) \to \infty $, which aligns with the well-known result that there is no true BEC transition in one dimension (Mermin-Wagner-Hohenberg theorem); instead, one observes quasi-condensation \cite{mermin1966, hohenberg1967, khawaja2003}. Finally, we note that for large Wigner parameters, the critical temperature becomes independent of $ \theta $, saturating at a maximum value of \begin{equation}
    2^{1-1/d} t_c(0). 
\end{equation}

\section{Thermodynamic Behavior in the Degenerate regime} \label{sec3}

The heat capacity can be readily determined from equation (\ref{uu}) using the well-known formula:
\begin{equation}
C = \frac{\partial U}{\partial T}.
\end{equation}
However, it is crucial to analyze the heat capacity in two distinct regimes:
\begin{itemize}
\item For temperatures below the transition point ($t < t_c$),
\item For temperatures above the transition point ($t > t_c$).
\end{itemize}
For temperatures below the transition temperature, we can safely set $z=1$, yielding:
\begin{equation}
\frac{C_{<}}{Nk_B}=
d(d+1) \frac{g_{d+1}(1,\theta)}{g_d(1,\theta)}\left(\frac{t}{t_c}\right)^d .
\label{cvi}
\end{equation}
The difference with the non-deformed case may be better visualized by rewriting this expression as follows:  
\begin{equation}
C_{<}(\theta)=
\frac{1}{2^d}\left(1+\frac{2^d-1}{(1+2\theta)^d}\right)C_{<}(\theta=0) ,
\label{ratio}
\end{equation}
which shows that, as expected, the standard $ t^d $ behavior of the heat capacity for a degenerate harmonically trapped ideal Bose gas remains unchanged despite the deformation of the underlying Heisenberg algebra. Furthermore, for sufficiently strong deformations, the heat capacity saturates to $ \frac{1}{2^d} $ of its non-deformed counterpart.

On the other hand, at temperatures above the transition, the fugacity becomes dependent on $ t $, which complicates the analysis. However, since $ N_0 \ll N_{\epsilon} \simeq N $ in this regime, Eq. (\ref{s}) simplifies, resulting in the following implicit relationship between $ z $ and \( t \):
\begin{equation}
\frac{g_d(z,\theta)}{g_d(1,\theta)}=\left(\frac{t_c}{t}\right)^d .
\label{z}
\end{equation}
From this, the heat capacity for $t>t_c$ reads:
\begin{equation}
\frac{C_{{>}}}{Nk_{B}}=d(d+1)\frac{
g_{d+1}(z,\theta )}{g_d (z,\theta) }-d^2\frac{
g_{d}(z,\theta )}{g_{d-1} (z,\theta) }.
\label{cvs}
\end{equation}
By computing the ratio \( C_>(\theta) / C_>(\theta = 0) \), we obtain the following expression:
\begin{equation}
\frac{C_{{>}}(\theta)}{C_{{>}}(0)}=
\frac{d(d+1)\dfrac{
g_{d+1}(1,\theta )}{g_d (1,\theta) }-d^2\dfrac{
g_{d}(1,\theta )}{g_{d-1} (1,\theta) }}
{d(d+1)\dfrac{g_{d+1}(1)}{g_d (1) }-d^2\dfrac{
g_{d}(1)}{g_{d-1} (1) }}
\left(\dfrac{t_c(\theta)}{t_c(0)}
\right)^{1/d} ,
\label{ratiocv}
\end{equation}
which is clearly independent of temperature. Therefore, \( C_> \) behaves like \( 1/t \), similar to the non-deformed case. 

Based on the previous calculations, we conclude that the overall temperature dependence of the heat capacity above and beyond the transition is unaffected by the Wigner parameter. Therefore, as expected from general symmetry considerations, the nature of the transition and the critical exponents remain unchanged.

\begin{figure}[hbtp]
\centering
\includegraphics[scale=1]{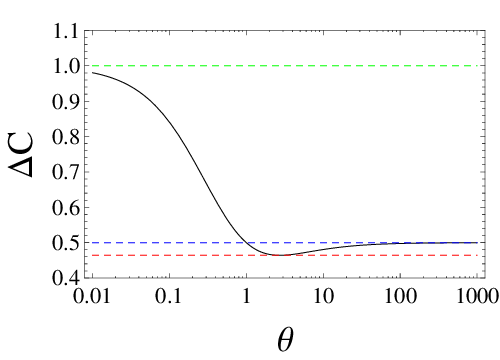}
\caption{Normalized discontinuity $\Delta C$ versus $\theta$ for $d=3$. The dashed lines (green, blue, red) represent the non-deformed case, the minimum value $2\sqrt{3}-3$, and the asymptotic value $1/2$, respectively.}
\label{fig1}
\end{figure}

Finally, it is worth mentioning that the Eqs. (\ref{cvi}) and (\ref{cvs}) for $C_<$ and $C_>$ are natural generalizations of the results obtained in \cite{gro1} (for d=2,3) for the non-deformed case. Moreover, in our previous work \cite{HO24}, we have also shown that, for $d=3$, the heat capacity exhibits typical $\lambda-$point behavior of a phase transition. The discontinuity at the critical point reads,
\begin{equation}
\Delta C=\frac{C_{<}-C_{>}}{Nk_{B}}\bigg|_{t_c}=
9\frac{\zeta(3)}{\zeta(2)}\frac{1+\theta+\theta^2}{(1+\theta)(1+2\theta)}
\label{jump}
\end{equation}
and reduces to the known value 
\begin{equation}
    54\zeta(3)/\pi^2 \simeq 6.577, \qquad \text{when} \qquad \theta = 0.
\end{equation}
As an illustration, we plot this jump (normalized to its value for $ \theta = 0 $) as a function of the Wigner deformation parameter in Figure \ref{fig1}. 

\newpage
\noindent For $-1/2<\theta<0$ (not represented in the figure), it grows to infinity, while for $\theta>0$, it decreases toward a minimum ($\simeq 0.464$) and then slightly grows again to reach the limiting value ($0.5$) for infinitely strong deformations. Since the discontinuity of the heat capacity can be measured experimentally \cite{cvexp1, cvexp2, cvexp3, cvexp4, cvexp5}, such departures from the standard value would witness a non-zero Wigner parameter. 

\begin{figure}[hbtp]
\centering
\includegraphics[scale=1]{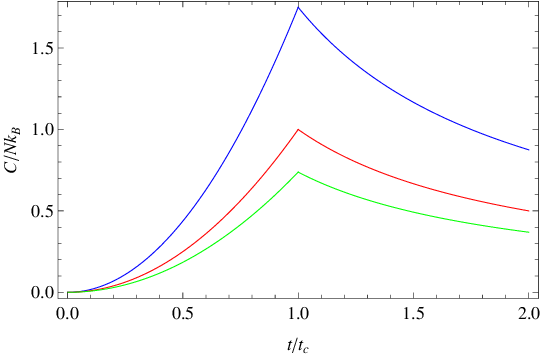}
\caption{Normalized heat capacity $C/Nk_B$ versus $t/t_c$ for $d=2$, for different values of the Wigner parameter. Green: $\theta=+0.2$, Red: $\theta=0$, Blue: $\theta=-0.2$.}
\label{fig2}
\end{figure}

For $ d = 2 $, the situation differs somewhat. Although the typical $ \lambda $-point behavior of a phase transition is still observed, we note from Eqs. (\ref{cvi}) and (\ref{cvs}) that the heat capacity remains continuous at the transition. This behavior is illustrated in Fig. \ref{fig2}, where we plot $ C/Nk_B $ normalized to its maximum value 
\begin{equation}
  6\zeta(3)/\pi^2, \qquad \text{for} \qquad \theta = 0.   
\end{equation}

\noindent As $ \theta $ decreases, the maximum increases and diverges as $ \theta \to -1/2 $. Thus, unlike the three-dimensional case, the phase transition remains continuous, resembling a Berezinskii–Kosterlitz–Thouless transition, yet it exhibits a sharp peak in the heat capacity.

\section{Classical regime and upper bound for the Wigner parameter} \label{sec4}

The classical, or high-temperature, regime is reached when $ t \gg t_c $. According to Eq. (\ref{z}), this implies $ g_d(z, \theta) \ll 1 $, which occurs only when $ z \ll 1 $. Expanding the generalized Bose functions around $ z = 0 $, we obtain:
\be\label{bosef}
g_d(z,\theta)\simeq \dfrac{ z^2}{2^{d-1}}+\dfrac{z^{1+2\theta}}{(1+2\theta)^{d-1}}+...
\ee
This expansion shows that the leading terms are governed by the deformation parameter. Consequently, for $ -\frac{1}{2} < \theta \leq \frac{1}{2} $, the internal energy and heat capacity functions are given by:
\be
\ba{rl}
U&=\dfrac{d}{1+2\theta}Nk_BT, \\
C_{>}&=\dfrac{d}{1+2\theta}Nk_B. \\
\ea
\ee
We observe that these values are greater (smaller) than those in the non-deformed case when $ -\frac{1}{2} < \theta < 0 $ ($ 0 < \theta < \frac{1}{2} $). Moreover, for $ \theta = 0 $, we recover the well-known results for an ideal gas confined in a $ d $-dimensional harmonic trap. The previous expressions remain valid as long as $ \theta < \frac{1}{2} $. The previous expressions remain valid as long as $ \theta < \frac{1}{2} $. However, at the critical value $ \theta = \frac{1}{2} $, Eq. (\ref{bosef}) reveals that the leading-order terms are given by $ z^2 $ and $ z^{1 + 2\theta} $, leading to the remarkable result:  
\be \label{classical}
\ba{rl}
U&=dNk_BT, \\
C_{>}&=dNk_B .\\
\ea
\ee
This demonstrates that the classical behaviors for $ \theta = 1/2 $ and $ \theta = 0 $ are essentially identical.

On the other hand, when $ \theta > 1/2 $, the leading terms of the expansions in Eq. (\ref{bosef}) become proportional to $ z^2 $, resulting in:
\be
\ba{rl}
U&=\dfrac{d}{2}\,Nk_BT, \\
C_{>}&=\dfrac{d}{2}\,Nk_B. \\
\ea
\ee
Surprisingly, these expressions are independent of \( \theta \) and are characteristic of the non-deformed ideal homogeneous classical gas, which is clearly not the system under consideration. Therefore, the only way to obtain the correct classical behavior is to restrict the Wigner parameter to the domain 
\begin{equation}
    -\frac{1}{2}< \theta \leq \frac{1}{2}.
\end{equation}
Thus, unlike previous studies where $ \theta$  was treated as a free deformation parameter with only a lower bound constraint, we demonstrate that this is not the case for a broad class of harmonic traps in different dimensions.

It is worth noting that the previous results can be easily extended to the ideal homogeneous Bose gas by replacing 
\be
\ba{rl}
d&\rightarrow \frac{d}{2}, \\
\hbar\omega &\rightarrow 2\pi\hbar^2/mV_d^{d/2},  \\
\ea
\ee
where $ V_d $ represents the generalized hypervolume. This extension is valid because, in both cases, the Hamiltonian is quadratic in its degrees of freedom. The number of nonvanishing coefficients for the quadratic terms is doubled when transitioning from the homogeneous case to the harmonically trapped case \cite{path2017}. Therefore, we can conclude that in the \( d \)-dimensional homogeneous Bose gas, the Wigner parameter must also be restricted to the same domain.

In conclusion, We would like to underscore that these bounds are crucial for the proper application of the Dunkl formalism. Specifically, if $ \theta $ is used as a fitting parameter, the only valid values must lie within this domain.

\section{Conclusions} \label{sec5}

In the Dunkl formalism, the Wigner parameter is typically treated as a free parameter. In this work, we have specifically examined its role in the context of a harmonically trapped ideal Bose gas in arbitrary $d$-dimensions. We found that, for reasons of consistency, this parameter must be greater than \(-1/2\). Additionally, we showed that this condition also corresponds to the requirement for the existence of a Bose-Einstein condensate.  

Moreover, by analyzing the high-temperature regime, we demonstrated that to maintain correct classical behavior, the Wigner parameter must be bounded from above as well, leading to the final constraint \(-1/2 < \theta \leq 1/2\). This result refines the physical applicability of the Dunkl formalism, ensuring that the deformation does not lead to unphysical thermodynamic predictions.  

The prediction of the presence of an upper bound in \(\theta\) has several important consequences. First, it guarantees thermodynamic consistency, preventing the system from exhibiting behavior inconsistent with its confinement. Second, it confirms that while \(\theta\) modifies thermodynamic functions, it does not alter the order of the phase transition or the critical exponents. Lastly, it provides a well-defined range for \(\theta\), which is crucial when using it as a fitting parameter in experimental or phenomenological studies.  

These results suggest natural extensions of the present work. A key direction for future research is the study of interacting Bose gases within the Dunkl framework. Since the Dunkl derivative effectively introduces an interaction term in the Hamiltonian, it would be valuable to explore whether the derived bounds for \(\theta\) persist in the presence of particle interactions. Another potential avenue is the investigation of the role of the Wigner parameter in non-equilibrium settings, such as quantum quenches or time-dependent perturbations. Additionally, the potential extension of this framework to more complex systems, particularly those involving spin degrees of freedom, could open new avenues for understanding the interplay between deformation and quantum statistics. The introduction of spin-dependent Dunkl derivatives could lead to novel symmetry-breaking effects and modifications in the thermodynamic properties of quantum gases. This direction may also provide insights into potential applications in quantum simulation, the study of exotic quantum phases, and the investigation of systems with fractional statistics.  

We believe that our findings provide an important step toward understanding the role of the Dunkl formalism in quantum statistical mechanics and look forward to future studies that build upon this work.

\newpage
\section*{Acknowledgments}
The authors thank the anonymous reviewers for their constructive comments and the editor for their valuable feedback and support. This work is supported by the Ministry of Higher Education and Scientific Research, Algeria under the code: PRFU:B00L02UN020120220002. B. C. L. is grateful to Excellence project PřF UHK 2205/2025-2026 for the financial support.

\section*{Data Availability Statements}

The authors declare that the data supporting the findings of this study are available within the article.

\section*{Declaration of Generative AI and AI-assisted technologies in the writing process}

The authors did not use any generative AI and AI-assisted technologies in the writing process.


\begin{thebibliography}{99}

% Differential-difference operators associated to reflection groups
\bibitem{Dunkl1} C. F. Dunkl,  \href{https://doi.org/10.1090/S0002-9947-1989-0951883-8}{T. Am. Math. Soc. \textbf{311}, 167 (1989).}

% One-dimensional quantum mechanics with Dunkl derivative
\bibitem{phys2} W. S. Chung, H. Hassanabadi, \href{https://doi.org/10.1142/S0217732319501906}{Mod. Phys. Lett. A \textbf{34}, 1950190 (2019).}

% New deformed Heisenberg algebra with reflection operator
\bibitem{Chung2021} W. S. Chung, H. Hassanabadi, \href{https://doi.org/10.1140/epjp/s13360-021-01186-5}{Eur. Phys. J. Plus \textbf{136}, 239 (2021).}

% Do the Equations of Motion Determine the Quantum Mechanical Commutation Relations?
\bibitem{Wigner1950} E. P. Wigner, \href{https://doi.org/10.1103/PhysRev.77.711}{Phys. Rev. \textbf{77}, 711 (1950).}


% Deformed Heisenberg algebra and fractional spin field in 2 + 1 dimensions
\bibitem{Plyushchay1994} M. S. Plyushchay, \href{https://doi.org/10.1016/0370-2693(94)90828-1}{Phys. Lett. B \textbf{320}, 91 (1994).}  

% Deformed Heisenberg Algebra, Fractional Spin Fields, and Supersymmetry without Fermions
\bibitem{Plyushchay1996} M. S. Plyushchay, \href{https://doi.org/10.1006/aphy.1996.0012}{Ann. Phys. \textbf{245}, 339 (1996).} 

% R deformed Heisenberg algebra
\bibitem{Miky1} M. S. Plyushchay, \href{https://doi.org/10.1142/S0217732396002927}{Mod. Phys. Lett. A \textbf{11}, 2953 (1996).} 


% Dunkl Operator Formalism for Quantum Many-Body Problems Associated with Classical Root Systems
\bibitem{Hikami1996} K. Hikami, \href{https://doi.org/10.1143/JPSJ.65.394}{J. Phys. Soc. Jpn. \textbf{65}, 394 (1996).}

% Deformed Heisenberg algebra with reflection
\bibitem{Plyushchay1997} M. S. Plyushchay, \href{https://doi.org/10.1016/S0550-3213(97)00065-5}{Nucl. Phys. B \textbf{491}, 619 (1997).} 

% R deformed Heisenberg algebra, anyons and d = (2+1) supersymmetry
\bibitem{Miky2} M. S. Plyushchay, \href{https://doi.org/10.1142/S0217732397001187}{Mod. Phys. Lett. A \textbf{12}, 1153 (1997).}


% Supersymmetry of parafermions
\bibitem{Miky3} S. Klishevich, M. S. Plyushchay, \href{https://doi.org/10.1142/S0217732399002881}{Mod. Phys. Lett. A \textbf{14}, 2739 (1999).}





%Three aspects of bosonized supersymmetry and linear differential field equation with reflection
\bibitem{Gamboa1999} J. Gamboa, M. S. Plyushchay, J. Zanelli, \href{https://doi.org/10.1016/S0550-3213(98)00832-3}{Nucl. Phys. B \textbf{543}, 447 (1999).} 


% HIDDEN NONLINEAR SUPERSYMMETRIES IN PURE PARABOSONIC SYSTEMS
\bibitem{Plyushchay2000} M. Plyushchay, \href{https://doi.org/10.1142/S0217751X00001981}{Int. J. Mod. Phys. A \textbf{15}, 3679 (2000).}

% Bosons, fermions and anyons in the plane, and supersymmetry
\bibitem{Horvathy2010} P. A. Hortv\'athy, M. Plyushchay, M. Valenzuela, \href{https://doi.org/10.1016/j.aop.2010.02.007}{Ann. Phys. \textbf{325}, 1931 (2010).}

% Fractional helicity, Lorentz symmetry breaking, compactification and anyons
\bibitem{Klishevich2001} S. M. Klishevich, M. Plyushchay, M. Rausch de Traubenberg, \href{https://doi.org/10.1016/S0550-3213%2801%2900442-4}{Nucl. Phys. B \textbf{616}, 419 (2001).}

% Anyon wave equations and the noncommutative plane
\bibitem{Horvathy2004} P. A. Horvathy, M. Plyushchay, \href{https://doi.org/10.1016/j.physletb.2004.05.043}{Phys. Lett. B \textbf{595}, 547 (2004).} 

% Common algebraic structure for the Calogero - Sutherland models
\bibitem{Kakei1996} S. Kakei, \href{https://doi.org/10.1088/0305-4470/29/24/002}{J. Phys. A: Math. Gen. \textbf{29}, L619 (1996).}


% Exact operator solution of the Calogero-Sutherland model
\bibitem{Lapointe1996} L. Lapointe, L. Vinet, \href{https://doi.org/10.1007/BF02099456}{Commun. Math. Phys. \textbf{178}, 425 (1996).} 



% On the hydrogen atom via the Wigner–Heisenberg algebra
\bibitem{Rodrigues2009} R. de Lima Rodrigues, \href{https://doi.org/10.1088/1751-8113/42/35/355213}{J. Phys. A \textbf{42}, 355213 (2009).}






% Dunkl graphene in constant magnetic field %
\bibitem{Bilel5} B. Hamil, B. C. L\"utf\"uo\u{g}lu, \href{https://doi.org/10.1140/epjp/s13360-022-03463-3}{Eur. Phys. J. Plus \textbf{137}, 1241 (2022).}

% The Dunkl oscillator in the plane: I. Superintegrability, separated wavefunctions and overlap coefficients
\bibitem{Genest20131} V. X. Genest, M. E. H. Ismail, L. Vinet, A. Zhedanov, \href{https://doi.org/10.1088/1751-8113/46/14/145201}{J. Phys. A: Math. Theor. \textbf{46}, 145201 (2013).}

% The singular and the 2:1 anisotropic Dunkl oscillators in the plane
\bibitem{Genest20132} V. X. Genest, L. Vinet, A. Zhedanov, \href{https://doi.org/10.1088/1751-8113/46/32/325201}{J. Phys. A: Math. Theor. \textbf{46}, 325201 (2013).} 

% The Dunkl Oscillator in the Plane II: Representations of the Symmetry Algebra
\bibitem{Genest20141} V. X. Genest, M. E. H. Ismail, L. Vinet, A. Zhedanov, \href{https://doi.org/10.1007/s00220-014-1915-2}{Commun. Math. Phys. \textbf{329}, 999 (2014).}


% Families of 2D superintegrable anisotropic Dunkl oscillators and algebraic derivation of their spectrum
\bibitem{Isaac2016} P. S. Isaac, I. Marquette, \href{https://doi.org/10.1088/1751-8113/49/11/115201}{J. Phys. A: Math. Theor. \textbf{49}, 115201 (2016).}

% Superintegrability of the Dunkl–Coulomb problem in three-dimensions
\bibitem{Ghazouani2020} S. Ghazouani, S. Insaf, \href{https://doi.org/10.1088/1751-8121/ab4a2d}{J. Phys. A: Math. Theor. \textbf{53}, 035202 (2020). }

% Superintegrability on the Dunkl oscillator model in three-dimensional spaces of constant curvature
\bibitem{Najafizade20223} S. H. Dong, A. Najafizade,  H. Panahi, W. S. Chung, H. Hassanabadi,  \href{https://doi.org/10.1016/j.aop.2022.169014}{Ann. Phys. \textbf{444}, 169014 (2022).}

% Information theoretic measures in one-dimensional Dunkl oscillator
\bibitem{Debraj2024} D. Nath, N. Ghosh, A. K. Roy, \href{https://doi.org/10.1063/5.0200405}{J. Math. Phys. \textbf{65}, 083511 (2024).}

% Superstatistics of the Dunkl oscillator
\bibitem{HH2021} H. Hassanabadi, M. de Montigny,
W. S. Chung, P. Sedaghatnia, \href{https://doi.org/10.1016/j.physa.2021.126154}{Physica A \textbf{580}, 126154 (2021).}

% The condensation of ideal Bose gas in a gravitational field in the framework of Dunkl-statistic %
\bibitem{phys5} F. Merabtine, B. Hamil, B. C. L\"{u}tf\"{u}o\u{g}lu, A. Hocine, M. Benarous, \href{https://doi.org/10.1088/1742-5468/acd106}{J. Stat. Mech. \textbf{5}, 053102 (2023).}

% The condensation of ideal Bose gas in a gravitational field in the framework of Dunkl-statistic %
\bibitem{Bilel3} B. Hamil, B. C. L\"utf\"uo\u{g}lu, \href{https://doi.org/10.1016/j.physa.2023.128841}{Physica A \textbf{623}, 128841 (2023).}

% On Dunkl-Bose-Einstein Condensation in Harmonic Traps
\bibitem{HO24} A. Hocine, B. Hamil, F. Merabtine, B. C. L\"{u}tf\"{u}o\u{g}lu, M. Benarous, \href{https://doi.org/10.31349/RevMexFis.70.051701}{Rev. Mex.  Fís. \textbf{70}, 051701 (2024).}

% Condensation of Ideal Dunkl-Bose Gas in Power-Law Traps %
\bibitem{phys6} A. Hocine, F. Merabtine, B. Hamil, B. C. L\"{u}tf\"{u}o\u{g}lu, M. Benarous, \href{https://doi.org/10.1007/s12648-024-03311-3}{Indian J. Phys. \textbf{99}, 775 (2024).}


% The Dunkl oscillator in three dimensions
\bibitem{Genest20142} V. X. Genest L. Vinet, A. Zhedanov, \href{https://doi.org/10.1088/1742-6596/512/1/012010}{J. Phys.: Conf. Ser. \textbf{512}, 012010 (2014).}

% SU(1,1) solution for the Dunkl oscillator in two dimensions and its coherent states
\bibitem{Salazar2017} M. Salazar-Ram\'irez, D. Ojeda-Guill\'en, R. D. Mota, V. D. Granados, \href{https://doi.org/10.1140/epjp/i2017-11314-3}{Eur. Phys. J. Plus \textbf{132}, 39 (2017).}

% Effect of the two-parameter generalized Dunkl derivative on the two-dimensional Schrödinger equation
\bibitem{Mota2022s} R. D. Mota, D. Ojeda-Guill\'{e}n, \href{https://doi.org/10.1142/S0217732322502248}{Mod. Phys. Lett. A \textbf{37},  2250224 (2022).}

% Generalized Dunkl-Schrodinger equations: solvable cases, point transformations, and position-dependent mass systems
\bibitem{Halberg2022} A. Schulze-Halberg, \href{https://doi.org/10.1088/1402-4896/ac807a}{Phys. Scr. \textbf{97}, 085213 (2022).}

% Relativistic solutions of generalized-Dunkl harmonic and anharmonic oscillators
\bibitem{Samira20222} S. Hassanabadi, J.K\v{r}\'{\i}\v{z}, B. C. L\"{u}tf\"{u}o\u{g}lu, H. Hassanabadi, \href{https://doi.org/10.1088/1402-4896/aca2f7}{Phys. Scr. \textbf{97}, 125305 (2022).}


%Exact solutions of the generalized Dunkl oscillator in the Cartesian system
\bibitem{Dong2023} S. H. Dong, L. F. Quezada, W. S. Chung, P. Sedaghatnia, H. Hassanabadi, \href{https://doi.org/10.1016/j.aop.2023.169259}{Ann. Phys. \textbf{451}, 169259 (2023).}

% Effect of the Wigner–Dunkl algebra on the Dirac equation and Dirac harmonic oscillator
\bibitem{Sargol2018} S. Sargolzaeipor, H. Hassanabadi, W. S. Chung, \href{https://doi.org/10.1142/S0217732318501468}{Mod. Phys. Lett. A \textbf{33}, 1850146 (2018).}

% Exact solution of the relativistic Dunkl oscillator in (2+1) dimensions
\bibitem{Mota20181} R. D. Mota, D. Ojeda-Guill\'{e}n, M. Salazar-Ram\'{\i}rez, V. D. Granados, \href{https://doi.org/10.1016/j.aop.2019.167964}{Ann. Phys. \textbf{411}, 167964 (2019).}

%  Algebraic approach for the one-dimensional Dirac–Dunkl oscillator
\bibitem{Ojeda2020} D. Ojeda-Guill\'{e}n, R. D. Mota, M. Salazar-Ram\'{\i}rez, V. D. Granados, \href{https://doi.org/10.1142/S0217732320502557}{Mod. Phys. Lett. A \textbf{35}, 2050255 (2020).}

% Landau levels for the (2 + 1) Dunkl–Klein–Gordon oscillator
\bibitem{Mota20211} R. D. Mota, D. Ojeda-Guill\'{e}n, M. Salazar-Ramirez, V. D. Granados, \href{https://doi.org/10.1142/S0217732321500668}{Mod. Phys. Lett. A \textbf{36},  2150066 (2021).}

% Exact solutions of the 2D Dunkl–Klein–Gordon equation: The Coulomb potential and the Klein–Gordon oscillator
\bibitem{Mota20212} R. D. Mota, D. Ojeda-Guill\'{e}n, M. Salazar-Ramirez, V. D. Granados, \href{https://doi.org/10.1142/S0217732321501716}{Mod. Phys. Lett. A \textbf{36},  2150171 (2021).} 

% The Dunkl–Duffin–Kemmer–Petiau Oscillator
\bibitem{Merad2021} A. Merad, M. Merad, \href{https://doi.org/10.1007/s00601-021-01683-4}{Few-Body Syst. \textbf{62}, 98 (2021).}


% Dunkl–Klein–Gordon Equation in Three-Dimensions: The Klein–Gordon Oscillator and Coulomb Potential
\bibitem{Bilel20221} B. Hamil, B. C. L\"{u}tf\"{u}o\u{g}lu, \href{https://doi.org/10.1007/s00601-022-01776-8}{Few-Body Syst. \textbf{63}, 74 (2022).}



% Dunkl–Duffin–Kemmer–Petiau equation in (2 + 1) dimensions: The Dunkl–Bosonic oscillator spinless under Landau levels and Coulomb potential
\bibitem{Merad2022} A. Merad, M. Merad, T. Boudjedaa, \href{https://doi.org/10.1142/S0217751X22500725}{Int. J. Mod. Phys. A \textbf{37}, 2250072 (2022).}

% DKP equation in Wigner–Dunkl quantum mechanics framework
\bibitem{Askari2023} A. Askari, H. Hassanabadi, W. S. Chung,  \href{https://doi.org/10.1142/S0217751X23500616}{ Int. J. Mod. Phys. A \textbf{38},  2350061 (2023).}




% Thermodynamics of boson systems related to Dunkl differential-difference operators
\bibitem{Ubriaco2014} M. R. Ubriaco, \href{https://doi.org/10.1016/j.physa.2014.06.087}{Physica A \textbf{414}, 128 (2014).}

% Exact solutions to generalized Dunkl oscillator and its thermodynamic properties
\bibitem{Dong2021} S. H. Dong, W. H. Huang, W. S. Chung, P. Sedaghatnia, H. Hassanabadi, \href{https://doi.org/10.1209/0295-5075/ac2453}{EPL \textbf{135}, 30006 (2021).}

% Thermal properties of relativistic Dunkl oscillators
\bibitem{Bilel20222} B. Hamil, B. C. L\"{u}tf\"{u}o\u{g}lu, \href{https://doi.org/10.1140/epjp/s13360-022-03055-1}{Eur. Phys. J. Plus \textbf{137}, 812 (2022).}

% The Klein-Gordon and Dirac oscillators with generalized Dunkl derivative
\bibitem{Rouabhia2023} N. Rouabhia, M. Merad, B. Hamil, \href{https://doi.org/10.1209/0295-5075/acf409}{EPL \textbf{143}, 52003 (2023).}


% Dunkl-Schr\"{o}dinger equation with time-dependent harmonic oscillator potential" }\\
\bibitem{Benchikha20241} A. Benchikha, B. Hamil,  B. C. L\"{u}tf\"{u}o\u{g}lu,  B. Khantoul, \href{https://doi.org/10.1007/s10773-024-05786-6}{Int. J. Theor. Phys.  \textbf{63}, 248 (2024).}


% Time-Dependent Dunkl-Schrödinger Equation with an Angular-Dependent Potential
\bibitem{Lutfuoglu2025} B. C. L\"{u}tf\"{u}o\u{g}lu,  A. Benchikha, B. Hamil, B. Khantoul, \href{https://doi.org/10.1142/S0217732325500099}{Mod. Phys. Lett. A  \textbf{40}, 2550009 (2025).} 

% The Dunkl–Coulomb problem in the plane
\bibitem{Vincent3} V. X. Genest, A. Lapointe, L. Vinet, \href{https://doi.org/10.1016/j.physleta.2015.01.023}{Phys. Lett. A \textbf{379}, 923 (2015).}



% SU(1,1) solution for the Dunkl–Coulomb problem in two dimensions and its coherent states
\bibitem{Salazar2018} M. Salazar-Ram\'irez, D. Ojeda-Guill\'en, R. D. Mota, V. D. Granados, \href{https://doi.org/10.1142/S0217732318501122}{Mod. Phys. Lett. A \textbf{33}, 1850112 (2018).}

% The Dunkl–Coulomb problem in three-dimensions: energy spectrum, wave functions and h-spherical harmonics
\bibitem{Ghazouani2019} S. Ghazouani, I. Sboui, M. A. Amdouni, M. B. El Hadj Rhouma \href{https://doi.org/10.1088/1751-8121/ab0d98}{J. Phys. A: Math. Theor. \textbf{52}, 225202 (2019).}

%Algebraic approach to the Dunkl–Coulomb problem and Dunkl oscillator in arbitrary dimensions
\bibitem{Ghazouani2021} S. Ghazouani, S. Insaf, \href{https://doi.org/10.1007/s13324-020-00470-4}{Anal. Math. Phys. \textbf{11}, 35 (2021).}



% Exact solution to two dimensional Dunkl harmonic oscillator in the Non-Commutative phase-space
\bibitem{Samira2022} S. Hassanabadi, P. Sedaghatnia, W. S. Chung, B. C. L\"utf\"uo\u{g}lu, J. K\u{r}\`i\u{z}, H. Hassanabadi, \href{https://doi.org/10.1140/epjp/s13360-023-03933-2}{Eur. Phys. J. Plus \textbf{138}, 331 (2023).}

% The anisotropic Dunkl oscillator problem on the two-dimensional curved spaces
\bibitem{Najafizade20221} A. Najafizade, H. Panahi, \href{https://doi.org/10.1142/S0217732322500237}{Mod. Phys. Lett. A \textbf{37}, 2250023 (2022).}

% A representation of the Dunkl oscillator model on curved spaces: Factorization approach
\bibitem{Najafizade20222} A. Najafizade,  H. Panahi, W. S. Chung,  H. Hassanabadi,  \href{https://doi.org/10.1063/5.0041830}{J. Math. Phys. \textbf{63}, 033505 (2022).}


% The Dunkl oscillator on a space of nonconstant curvature: An exactly solvable quantum model with reflections
\bibitem{Ballesteros2023} A. Ballesteros, A. Najafizade, H. Panahi, H. Hassanabadi, S. H. Dong, \href{https://doi.org/10.1016/j.aop.2023.169543}{Ann. Phys. \textbf{460}, 169543 (2024).}

% On the gauge invariance of Wigner–Dunkl quantum mechanics in the presence of a constant magnetic field
\bibitem{Junker2023} G. Junker, S. H. Dong, P. Sedaghatnia, W. S. Chung, H. Hassanabadi, \href{https://doi.org/10.1016/j.aop.2023.169336}{Ann. Phys. \textbf{454}, 169336 (2023).}

% Quasi-exactly solvable potentials in Wigner-Dunkl quantum mechanics
\bibitem{Quesne2024} C. Quesne, \href{https://doi.org/10.1209/0295-5075/ad2947}{EPL \textbf{145}, 62001 (2024).}

% Dunkl–Pauli equation in the presence of a magnetic field.
\bibitem{Bougerne}H. Bouguerne, B. Hamil, B. C. L\"{u}tf\"{u}o\u{g}lu, M. Merad,
\href{ https://doi.org/10.1007/s12648-024-03170-y}{Indian J. Phys. \textbf{98}, 4093 (2024).}  

% Dunkl Algebra and Vacuum Pair Creation: Exact Analytical Results via Bogoliubov Method%
\bibitem{Bouguerne20243} H. Bouguerne, B. Hamil, B. C. L\"{u}tf\"{u}o\u{g}lu, M. Merad,
\href{https://doi.org/10.1016/j.nuclphysb.2024.116684}{ Nucl. Phys. B \textbf{1007}, 116684 (2024).}

% On the path integral formulation of Wigner–Dunkl quantum mechanics
\bibitem{Junker20241} G. Junker, \href{https://doi.org/10.1088/1751-8121/ad213d}{J. Phys. A: Math. Theor. \textbf{57}, 075201 (2024).}

% Path integral formulation for Dunkl-Dirac oscillator in (1+1)-dimensional space-time coordinates
\bibitem{Benzair2024} H. Benzair, T. Boudjedaa, M. Merad, \href{https://doi.org/10.1088/1402-4896/ad39b7}{Phys. Scr. \textbf{99}, 055261 (2024).}



% One-dimensional Dunkl Quantum Mechanics: A Path Integral Approach
\bibitem{Benchikha20242} A. Benchikha, B. Hamil,  B. C. L\"{u}tf\"{u}o\u{g}lu,  B. Khantoul, \href{https://doi.org/10.1088/1402-4896/ad7aaf}{Phys. Scr.   \textbf{99}, 105274 (2024).}

% Closed-form solutions of the Dunkl–Klein–Gordon equation for two inverse power-law interactions
\bibitem{Schulze20245} A. Schulze-Halberg, \href{https://doi.org/10.1142/S0217732323501134}{Int. J. Mod. Phys. A \textbf{39}, 2450013 (2024).}

% Approximate bound state solutions of the Dunkl-Schrödinger equation for a hyperbolic double-well interaction
\bibitem{Schulze20243} A. Schulze-Halberg, \href{https://doi.org/10.1088/1402-4896/ad4fc8}{Phys. Scr. \textbf{99}, 075212 (2024).}

% Approximate Solutions of the Dunkl–Schrödinger Equation for the Hyperbolic Pöschl–Teller Potential
\bibitem{Schulze20244} A. Schulze-Halberg, \href{https://doi.org/10.1007/s00601-024-01931-3}{Few-Body Syst. \textbf{65}, 58 (2024).}

% The generalized Fokker–Planck equation in terms of Dunkl-type derivatives
\bibitem{Mota20241} R. D. Mota, D. Ojeda-Guill\'{e}n,  M. A. Xicot\'encatl, \href{https://doi.org/10.1016/j.physa.2024.129525}{Physica A \textbf{635},  129525 (2024).}

% The Dunkl–Fokker–Planck Equation in (1+1) Dimensions
\bibitem{Mota20242} R. D. Mota, D. Ojeda-Guill\'{e}n,  M. A. Xicot\'encatl,  \href{https://doi.org/10.1007/s00601-024-01898-1}{Few-Body Syst. \textbf{65}, 25 (2024).}




% Dunkl-Schr\"odinger Equation in Higher Dimensions" 
\bibitem{Hamil20251} B. Hamil, B. C. L\"{u}tf\"{u}o\u{g}lu, M. Merad, \href{https://doi.org/10.1088/1402-4896/ada9b2}{Phys. Scrp.  \textbf{100},  035301 (2025).}

% A Path Integral Treatment of Time-dependent Dunkl Quantum Mechanics" }\\
\bibitem{Benchikha2025} A. Benchikha, B. Hamil, B. C. L\"{u}tf\"{u}o\u{g}lu, \href{https://doi.org/10.1142/S0219887825501130}{Int. J. Geom. Method. Mod. Phys. \textbf{XX}, XXXX (2025).}

% GUP to all Orders in the Planck Length: Some Applications. 
\bibitem{hamil2022} B. Hamil, B. C. Lütfüoğlu, \href{https://doi.org/10.1007/s10773-022-05188-6}{Int. J. Theor. Phys. \textbf{61}, 202 (2022).}

% Effect of the minimal length uncertainty relation on the density of states and the cosmological constant problem
\bibitem{chang2002} L. N. Chang, D. Minic, N. Okamura, T. Takeuchi, \href{https://doi.org/10.1103/PhysRevD.65.125028}{Phys. Rev. D \textbf{65}, 125028 (2002).}


% Rigorous Hamiltonian and Lagrangian analysis of classical and quantum theories with minimal length
\bibitem{bosso2018} P. Bosso, \href{https://doi.org/10.1103/PhysRevD.97.126010}{Phys. Rev. D \textbf{97}, 126010 (2018).}







% Bose–Einstein condensation in the framework of κ-statistics
\bibitem{kanidakis2003} A Aliano, G Kaniadakis, E Miraldi, \href{https://doi.org/10.1016/S0921-4526(02)01425-4}{Phys. B: Condens. Matter \textbf{325}, 35 (2003).}  

% Nonextensive boson gas and specific heat of 4He superfluid
\bibitem{wang2000} Q. A. Wang, M. Pezeril, A. Le M\'ehaut\'e, \href{https://doi.org/10.1016/S0378-4371(99)00639-1}{Physica A \textbf{278}, 337 (2000).}

% On Bose-Einstein condensation in harmonic traps
\bibitem{gro1} S. Grossmann, M. Holthaus, \href{https://doi.org/10.1016/0375-9601(95)00766-V}{Phys. Lett. A \textbf{208}, 188 (1995).}


% Absence of Ferromagnetism or Antiferromagnetism in One- or Two-Dimensional Isotropic Heisenberg Models
\bibitem{mermin1966} N. D. Mermin, H. Wagner, \href{https://doi.org/10.1103/PhysRevLett.17.1133}{Phys. Rev. Lett. \textbf{17}, 1133 (1966). }\emph{Erratum Phys. Rev. Lett. \textbf{17}, 1307 (1966).}

% Existence of Long-Range Order in One and Two Dimensions
\bibitem{hohenberg1967} P. C. Hohenberg, \href{https://doi.org/10.1103/PhysRev.158.383}{Phys. Rev. \textbf{158}, 383 (1967).}

% Dimensional and temperature crossover in trapped Bose gases
\bibitem{khawaja2003} U. Al Khawaja, N. P. Proukakis, J. O. Andersen, M. W. J. Romans,  H. T. C. Stoof,  \href{https://doi.org/10.1103/PhysRevA.68.043603}{Phys. Rev. A \textbf{68}, 043603 (2003).}

% A High Precision Scanning Ratio Calormimeter for use Near Phase Transitions
\bibitem{cvexp1} M. J. Buckingham, C. Edwards, J. A. Lipa, \href{https://doi.org/10.1063/1.1686346}
{Rev. Sci. Instrum. \textbf{44}, 1167 (1973).}

% Technique for determination of accurate heat capacities of volatile, powdered, or air-sensitive samples using relaxation calorimetry 
\bibitem{cvexp2} 
R. A. Marriott, M. Stancescu, C. A. Kennedy,  M. A. White, \href{https://doi.org/10.1063/1.2349606}{Rev. Sci. Instrum. \textbf{77}, 096108 (2006).}

%  Calorimetry of a Bose–Einstein-condensed photon gas,photon gas.  
\bibitem{cvexp3}
T. Damm, J. Schmitt, Q. Liang, D. Dung, F. Vewinger, M. Weitz, J. Klaers, \href{https://doi.org/10.1038/ncomms11340}{Nat. Commun. \textbf{7}, 11340 (2016).}

\bibitem{cvexp4}
Z. C. Tan, Q. Shi, Z. D. Nan, Y. Y. Di, \emph{Enthalpy and Internal Energy: Liquids, Solutions and Vapours,} ed. E. Wilhelm and T. Letcher, \href{https://doi.org/10.1039/9781788010214}{(The Royal Society of Chemistry, 2017), Chp. 23, pp. 590-610.}

% Standard methods for heat capacity measurements on a Quantum Design Physical Property Measurement System
\bibitem{cvexp5} P. F. Rosen,  B. F. Woodfield, \href{https://doi.org/10.1016/j.jct.2019.105974}{J. Chem. Thermodyn. \textbf{141}, 105974 (2020).}.

\bibitem{path2017} R. K. Pathria,  \emph{Statistical Mechanics: International Series of Monographs in Natural Philosophy} \href{https://www.perlego.com/book/1883802/statistical-mechanics-international-series-of-monographs-in-natural-philosophy-pdf}{(Elsevier,  2017). Vol. 45.} 





\end{thebibliography}
\end{document}